\documentclass{optica-article}

\journal{opticajournal} 

\articletype{Research Article}

\usepackage{lineno}
\usepackage{parskip} 
\usepackage{graphicx} 
\usepackage{float}
\usepackage{subcaption}

\begin{document}
\title{Towards Speaker Identification with Minimal Dataset and Constrained Resources using 1D-Convolution Neural Network}

\author{Irfan Nafiz Shahan\authormark{1,*}, Pulok Ahmed Auvi\authormark{1}}

\address{\authormark{1}Department of Electrical and Electronics Engineering, Shahjalal University of Science and Technology}


\begin{abstract*} 
Voice recognition and speaker identification are vital for applications in security and personal assistants. This paper presents a lightweight 1D-Convolutional Neural Network (1D-CNN) designed to perform speaker identification on minimal datasets. Our approach achieves a validation accuracy of 97.87\%, leveraging data augmentation techniques to handle background noise and limited training samples. Future improvements include testing on larger datasets and integrating transfer learning methods to enhance generalizability. We provide all code, the custom dataset, and the trained models to facilitate reproducibility. These resources are available on our GitHub repository: https://github.com/IrfanNafiz/RecMe.
\end{abstract*}

\begin{figure}[ht!]
\centering\includegraphics[width=\linewidth]{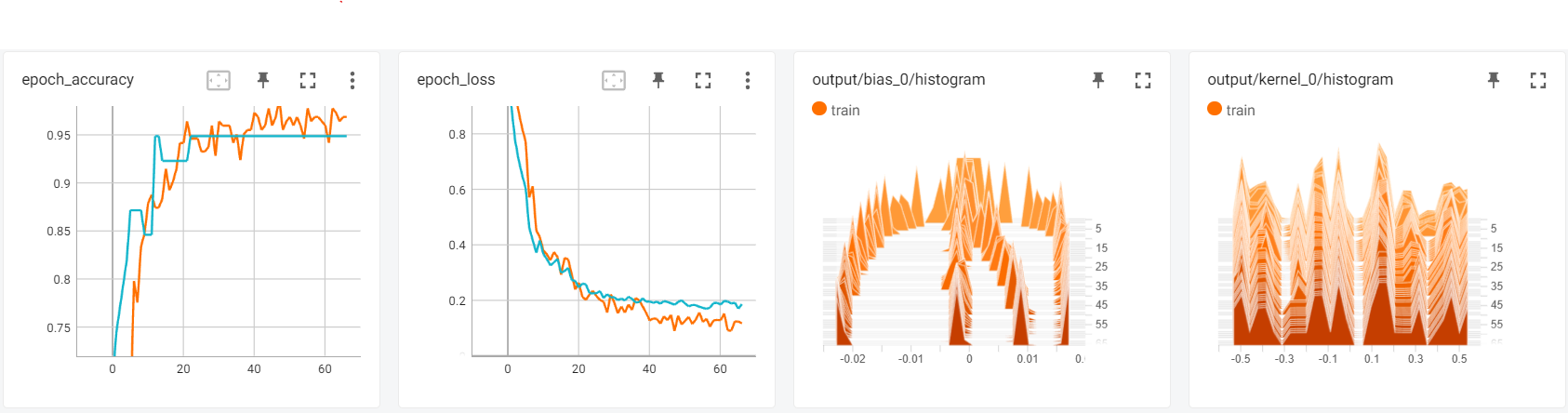}
\caption{1d-ConvNet Accuracy, Loss, and Output Layer Metrics}
\end{figure}

\section{Introduction}
Speaker identification (SI) is the task of recognizing the identity of someone based on the speaker’s speech signal. It is an important bio-feature recognition method that has many applications in security, forensics, and biometrics. One of the challenges of speaker identification is to extract robust and discriminative features from speech signals that can capture the speaker-specific characteristics. Recently, deep neural networks (DNNs) have been widely used for speaker identification, as they can learn high-level representations from raw or low-level features. In this paragraph, we will briefly review some of the current works in speaker identification using DNNs.

In a society reliant on voice interactions, accurately recognizing individuals through their unique vocal patterns has become a pressing need. Our project focuses on developing an innovative SI system that leverages DNNs to differentiate between speakers.

Despite recent advances in deep learning for speaker identification, the challenge of working with small, domain-specific datasets persists. Pretrained models like wav2vec and x-vectors have demonstrated success in this domain, but their computational complexity limits applicability in resource-constrained environments. We utilize a 1-Dimensional Convolutional Neural Network (1d-ConvNet) architecture, tailored to process voice data efficiently and effectively even in the face of small datasets and resource constraints.

In this work, we bridge this gap by developing a robust 1D-CNN framework tailored for small datasets, incorporating efficient data preprocessing and augmentation. This approach balances performance and computational efficiency. 

\section{Literature Review}

One of the popular DNN models for speaker identification is based on convolutional neural networks (CNNs), which can exploit the spatial features of voiceprints (corresponding to the voice spectrum) from spectrograms or mel-filterbank energy features (MFCCs). For example, \cite{ye2021deep} proposed a DNN model based on a two-dimensional CNN (2-D CNN) and gated recurrent unit (GRU) for SI. The 2-D CNN layer was used for voiceprint feature extraction and dimensionality reduction, while the stacked GRU layers were used for frame-level feature extraction. The model achieved a high recognition accuracy of 98.96\% on the Aishell-1 speech dataset.\cite{AIshallspeak} Recent models like wav2vec and HuBERT utilize self-supervised learning to extract high-level speech features from unlabeled data, enabling transfer learning for downstream tasks like speaker recognition \cite{hubert, schneider2019wav2vec}. While effective, these models require extensive computational resources and large-scale pretraining datasets.

Another popular DNN model for speaker identification is based on recurrent neural networks (RNNs), which can model the temporal dependencies of speech signals. RNNs can be further enhanced by using long short-term memory (LSTM) or GRU cells, which can overcome the vanishing gradient problem and capture long-term dependencies. For example, \cite{zhang2019text} proposed a DNN model based on LSTM and attention mechanism for speaker identification. The LSTM layer was used to encode the sequential information of speech signals, while the attention layer was used to weight the importance of each frame. The model achieved a recognition accuracy of 97.7\% on the VoxCeleb1 dataset. \cite{Nagrani17}

A recent trend in speaker identification is to use self-supervised learning methods, which can learn representations from unlabeled data by solving a pretext task. For example, \cite{chung2021hubert} proposed a DNN model based on HuBERT \cite{hsu2021hubert}, which is a transformer-based model that learns from masked acoustic features. The model was pre-trained on a large-scale unlabeled dataset and fine-tuned on a downstream speaker identification task. The model achieved state-of-the-art results on several benchmarks, such as VoxCeleb1 and VoxCeleb2.

Our work diverges from these high-resource approaches by focusing on optimizing small datasets with lightweight architectures. This includes incorporating augmentation techniques like noise addition and pitch shifting, commonly seen in resource-constrained setups \cite{Bousquet2020}.

\section{Methodology}
The entire program methodology has a few key steps:
\begin{enumerate}
    \item Data Collection
    \item Data Preprocessing
    \item Data Organization for Training
    \item Training the Deep Learning Model
    \item Application Decision-making 
\end{enumerate}

\subsection{\textbf{Data Collection}}
First, we record the voices of the speaker we want the model to recognize. In our program, we recorded almost 1 minute of our team members saying "Hello D S P 1 2 3 4 5". The recording files were named according to the speaker and saved in .wav format labeling the file name as per given by the speaker respectively.

\begin{figure}[h]
    \centering
    \includegraphics[width=0.75\textwidth]{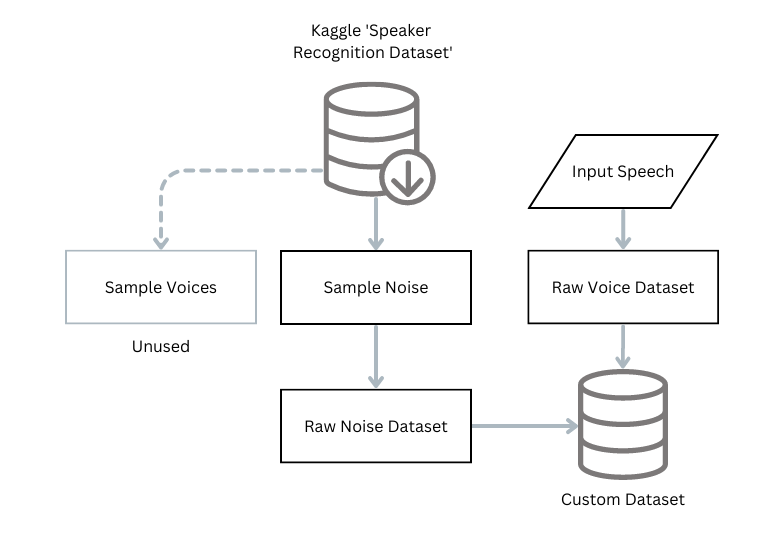}
    \caption{Creation of Raw Audio Dataset and Raw Noise Datasets to be used.}
    \label{fig:raw_dataset_creation}
\end{figure}

Moreover, some unique samples of background noise were collected from Keras' publicly available 'Speaker Recognition Dataset".\cite{SpeakerRecognitionDataset} There were a total of 6 different noise samples. See Fig.\ref{fig:raw_dataset_creation}.

\newpage
\subsection{\textbf{Data Preprocessing}}
Then, we resample each of the recordings and noises at 16,000Hz (an industry standard) and turn each of the recordings into 1-second clips. This creates about 60 clips for each person's dataset. This process of resampling and clipping is the preprocessing step and is necessary for the Deep Machine Learning Algorithm. A flow chart is shown in Fig.\ref{fig:preprocessing_database}.

Preprocesing step involves:
\begin{enumerate}
    \item Resampling at 16KHz because the sampling rate directly correlates to the number of frequency-bins in the Fourier Transform used to train our 1dConvNet DNN Algorithm.
    \item Clipping the 1 minute file into 60 1-second file provides a greater number of inputs for the ML Model to train on.
\end{enumerate}

\subsection{\textbf{Data Organization for Training}}
The recorded and preprocessed samples are then converted to a TensorFlow dataset, which contains a unique key for each user, and also the path to the preprocessed clips. Around 80\% of the dataset is used to train the model, knows as the training set, and the remaining 20\% will serve as test (or validation) set to monitor the model performances and tune it's hyperparameters. See Fig.\ref{fig:training_flow}.

Same is done for the noise clips.

\begin{figure}[ht!]
    \centering
    \includegraphics[width=0.75\linewidth]{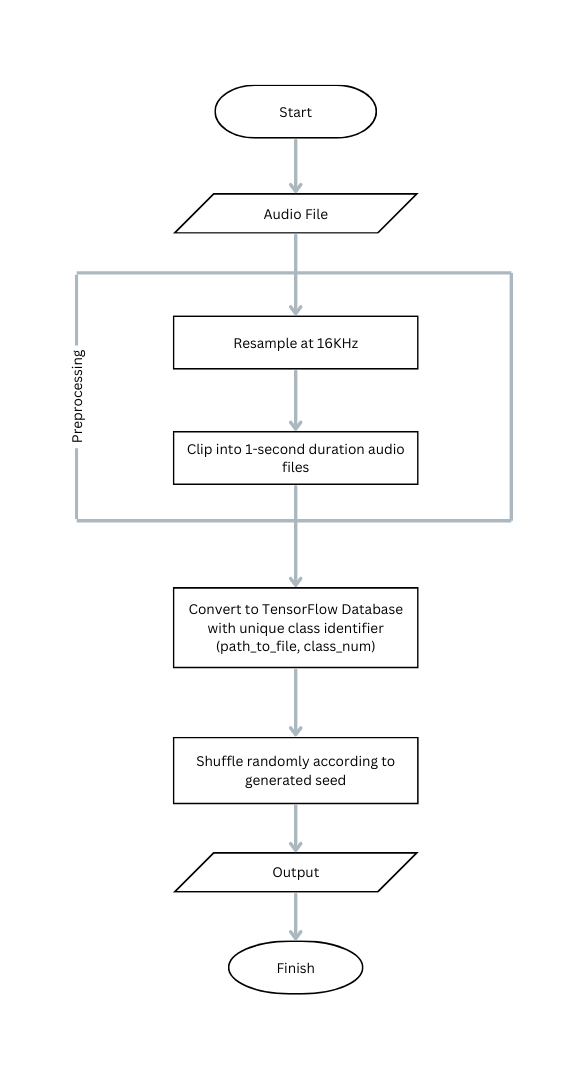}
    \caption{Pre-processing and Database Generation Flow-Chart}
    \label{fig:preprocessing_database}
\end{figure}

The files are then shuffled on their way to the training model so that it can learn to recognize the speaker based on their voice profiles and not by what they were saying. This allows the model to recognize the speaker even when the speaker is saying something other than "Hello D S P 1 2 3 4 5" in our case. See Fig.\ref{fig:preprocessing_database}.

Although this is a naive method, we show that the model works remarkably well even for unknown phrases!

Furthermore, the preprocessed noise sample is added to the training set and the test set so the model can learn to predict the speaker with an acceptable level of accuracy even with any atmospheric noise present.

\newpage

\subsection{\textbf{Training the Deep Learning Model}}
The Fast Fourier Transform (FFT) coefficients of each set of training and validation, are sent to the model inputs. 

The training was initiated using TensorFlow GPU on hardware specifications of Nvidia RTX 3060 6GB GPU built on top of an AMD Ryzen-7 5800H 16 Core CPU and 32 GB of RAM.

As the deep learning (DL) model trains, it compares itself with the training set and validation set (which it has not seen) according to the "Sparse Categorical Cross-Entropy" loss function and gains an estimate of how well it is performing both in the training and validation sets. This provides us with accuracy and loss metrics that are further details in Section.\ref{sec:eval_metrics}.

We are most concerned about maximizing validation accuracy because that metric tells us how well the model performs on unseen data, in our case, that is the speaker's voice input into the model. Since the DL problem is related to multiple classification, we want to minimize the validation loss values, but it is not our primary concern. 

Sparse Categorical Cross Entropy is used because it is an effective loss function for multiple classification DL scenarios, which is true in our case.

The goal here is to have a peak level of training and validation accuracy because it signifies that the model has learned to recognize the speakers well. But also the metrics for accuracy and loss should be healthy to ensure that the model indeed converges to a classification solution. Further details are provided in Section.\ref{sec:eval_metrics} 

\begin{figure}
    \centering
    \includegraphics[width=0.85\linewidth]{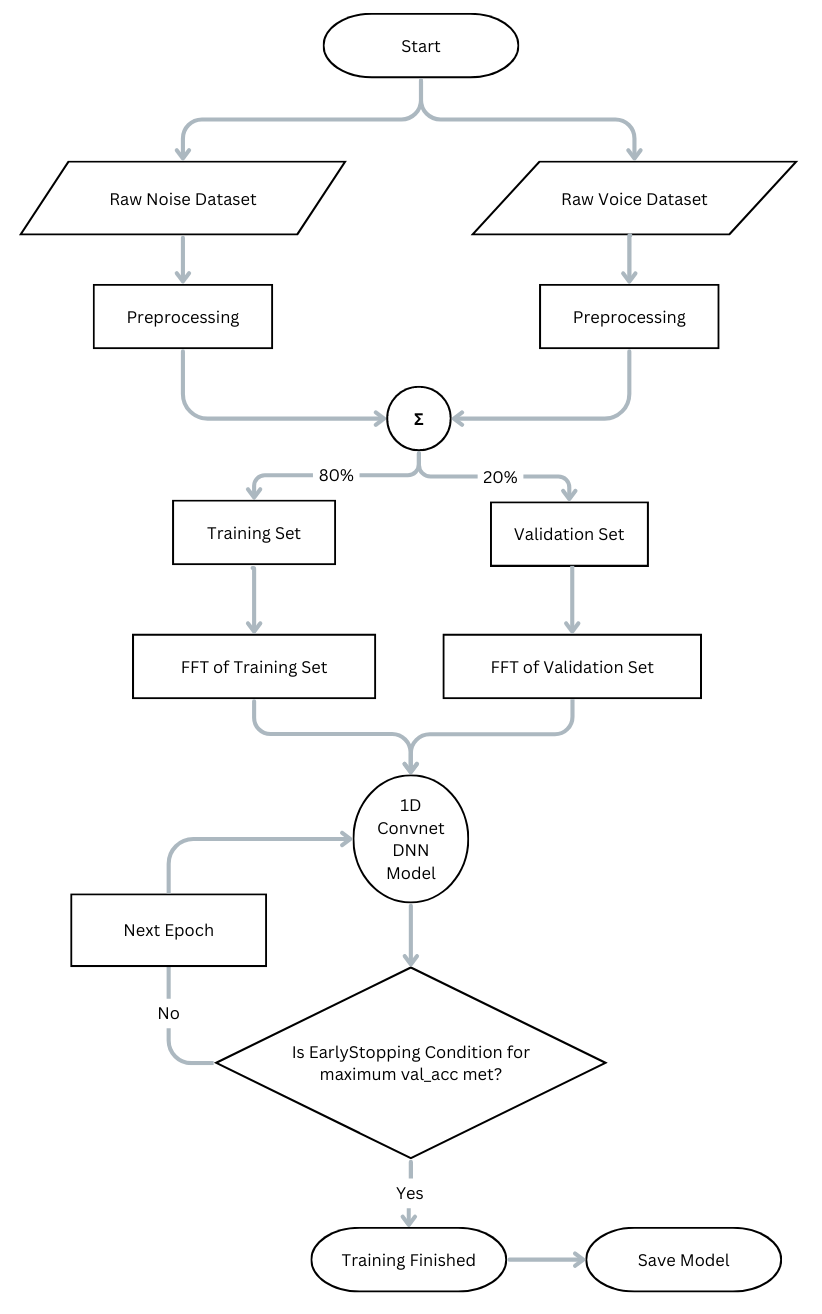}
    \caption{Training Protocol for the DL Model}
    \label{fig:training_flow}
\end{figure}

To ensure that the validation accuracy is maximized, some tuning is performed to determine the model architecture, number of neurons, and also other model hyperparameters. Furthermore to ensure model performance does not fall, we use an early stopping algorithm to stop our training when necessary. Moreover, to ensure proper convergence to the maximum possible validation accuracy is acquired, we use a learning rate scheduler. See Section.\ref{sec:ml} for more details on model architecture and parameters. A comprehensive flow chart for the training protocol is given in Fig.\ref{fig:training_flow}.

Once the model is fully trained according to the conditions mentioned above, the model parameters are saved as an h5 file, which is then loaded by our application during the speaker prediction process, hence a "Pretrained model" is used for prediction, leading to more efficient performance, instead of training again which would be performance and cost intensive. See Fig.\ref{fig:prediction_flow} to observe the use of pretrained model.

\newpage
\newpage

\subsection{\textbf{Application Decision-making}}
\begin{figure}[h]
    \centering
    \includegraphics[width=0.95\linewidth]{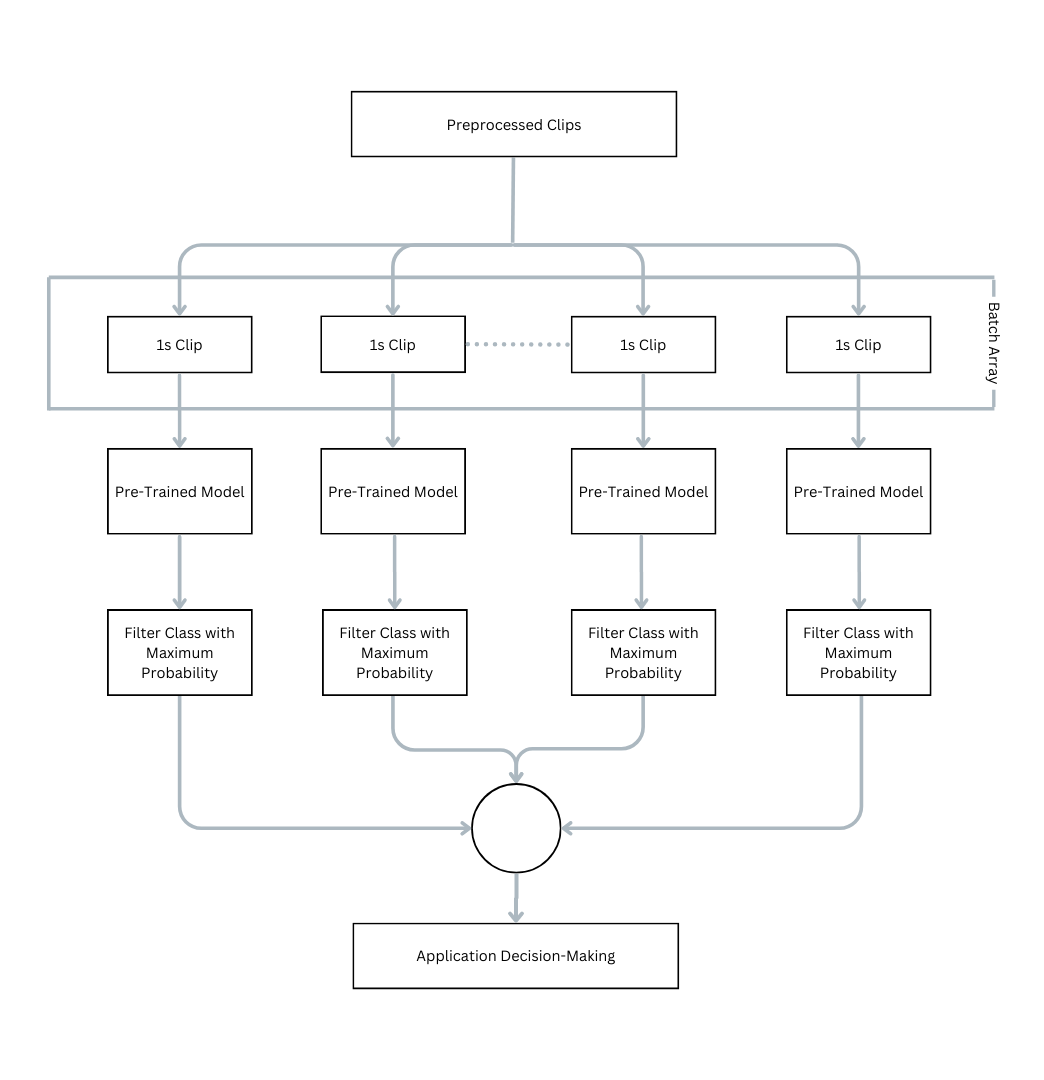}
    \caption{The prediction protocol flow-chart of the Model used in decision-making}
    \label{fig:prediction_flow}
\end{figure}

After training the model, we can test how well our model performs by using it to predict the speaker given proper inputs. Prompt the program to take any live voice recording as input. This recorded file is preprocessed again using the previous protocol into 1 sec clips, and each of the 1-second duration clips is fed to the pre-trained model in batches to calculate the probability of this voice belonging to any of the speaker it knows. See Fig.\ref{fig:prediction_flow}

A prediction is made for each 1s audio clip. The speaker that is predicted most for our recorded sample of clips is returned as the final output prediction.

\begin{figure}[ht!]
    \centering
    \includegraphics[width=0.6\linewidth]{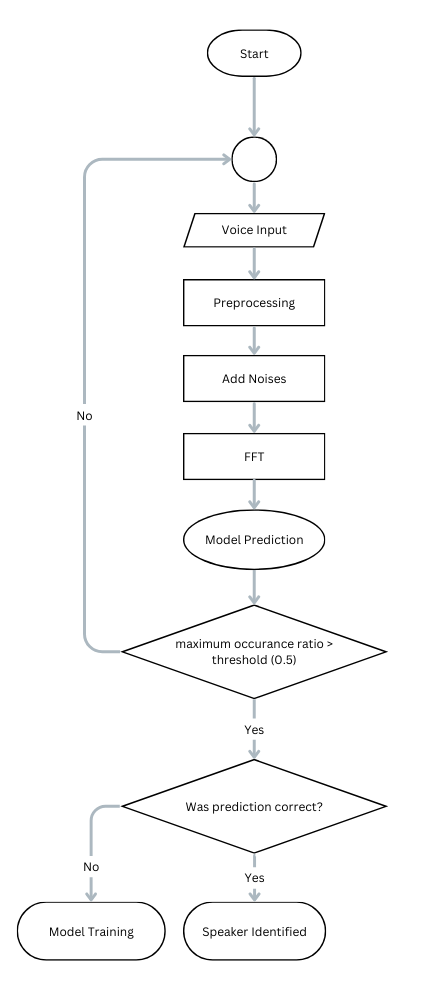}
    \caption{The decision-making flowchart for the application }
    \label{fig:decisionmaking_flow}
\end{figure}

If the test sample comes from a person that the model has not trained on, the highest probability of match will be very low. In that case, the model will say it does not recognize the test sample with enough confidence and ask for verification if the user is identified properly, if not, it will prompt the new user to make a new dataset for the new speaker. The model will then retrain itself to be able to classify each of the previous and new speakers it has data on. See Fig.\ref{fig:decisionmaking_flow} for a complete flow chart. Otherwise, the speaker is correctly identified and the application loop continues.

\newpage
\section{1D-Conv DNN Model - Taking a deeper look}
\label{sec:ml}
\subsection{Model Architecture}
\begin{figure}[h]
    \centering
    \includegraphics[width=0.75\linewidth]{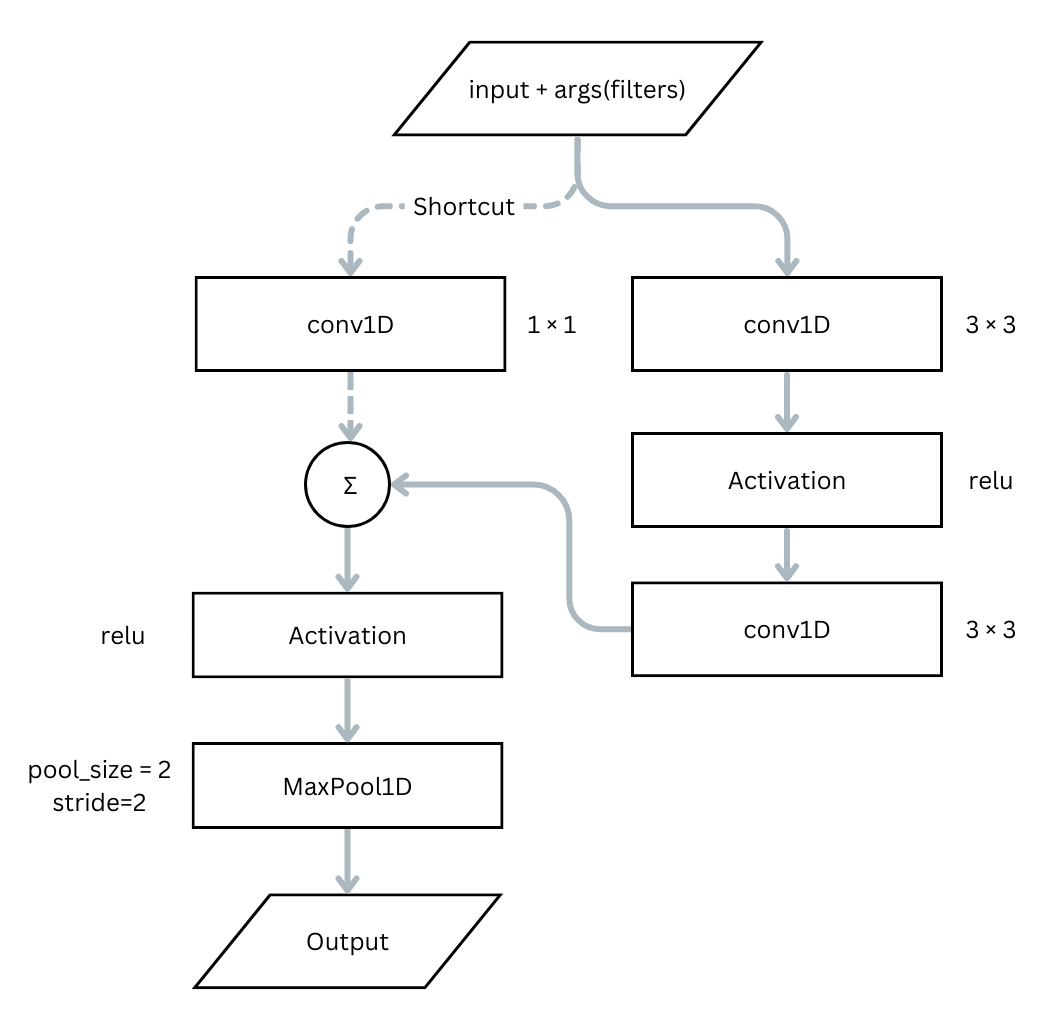}
    \caption{A flow chart of the residual block defined and called in our model architecture}
    \label{fig:res_block_flow}
\end{figure}

The machine learning model's core functionality is the use of 1D-Convolutional DNN layers. The comprehensive example used to build upon our model is the Keras Speaker Recognition Example publicly available online \cite{KerasSpeakerRecognitionExample}. The example was a great starting point, in order to assess the DL problem, and laid a foundation in order to train on smaller, manageable datasets for out application case with high accuracy after tuning and optimizing. The convolutional layers each are integrated into blocks, called "Residual Blocks" detailed in Section.\ref{subsubsec:res_block}. The flow chart of the residual block are given in Fig.\ref{fig:res_block_flow}.

\begin{figure}[h]
    \centering
    \includegraphics[width=0.95\linewidth]{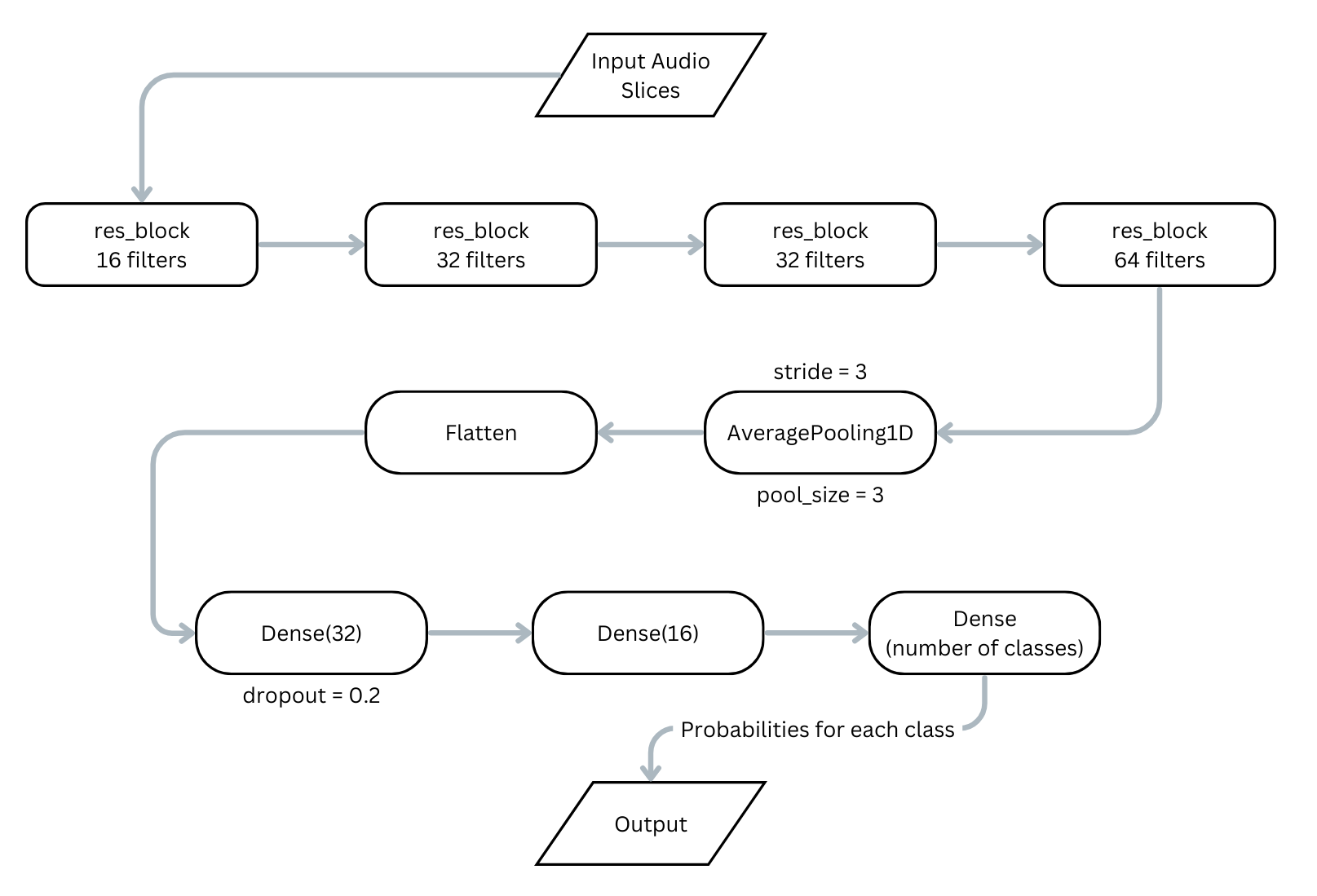}
    \caption{Complete model architecture flowchart}
    \label{fig:model_architecture_flow}
\end{figure}

The 4 residual blocks are concatenated according to performance of the model with filter parameters sent to them into a residual block series, and their final output is fed to an AveragePooling1D layer. This layer further reduces the dimensions of the output matrix from the residual block series. Afterwards the signal is sent to a series of 3 dense layers after passing through a Flatten layer to reduce the number of dimensions in out output matrix. For each dense layer number of neuron hyperparameters were tuned according to the performance of the model, with the first largest dense layer having a dropout regularization of 0.2, before being passed on the proceeding layers. A detailed flowchart of the model architecture overview is given in Fig.\ref{fig:model_architecture_flow}

The final layer is a dense layer with number of neurons equal to the number of classes in our model and a Softmax activation function. Softmax activation is necessary parameter in the final layers of multiple classification DL problems. As the number of speakers increase, so does the size of the final layer. This is done dynamically during our training.

\newpage
\subsubsection{Residual Block}
\label{subsubsec:res_block}
The residual block defined, takes number of filters as arguments and passes them onto each conv1D layer. The number of filters defines how many discrete samples are convoluted together within the kernel of the conv1D layers. The kernel sizes are 3x3 for the main branch and 1x1 for the shortcut branch. Refer to Fig.\ref{fig:res_block_flow}.

The shortcut branch that takes the 1x1 Conv1D of the input to the residual layer essentially mitigates the vanishing gradient problem often faced in complex DL networks when convolutional layers are involved. vanishing gradient problem is an issue during backpropagation algorithm that takes place during training, where the training parameters are defined by the gradient-descent algorithm to minimize the loss function of the model. This is a common practice in well known DL algorithms such as Residual Networks (ResNet) that perform similar calculation. Although ResNet is infamously taxing and requires significantly more time to train on typical models.

In the main path, the input to the block is passed through 2 convolution layers and an activation layer in between. 

The output of the shortcut path is element-wise added to the output of the main path afterwards which mitigates the vanishing gradient problem mentioned earlier. An activation and MaxPool1D layer is then incorporated to get the outputs from the training performed previously in the block, and sent to the output.

In all cases, Rectified Linear Units (ReLu) activation is taken as it is a standard for most DL cases. 

The residual block is paramount in capturing important features within the FFT of our processed samples that refer to each speaker. This enables the model to train particularly on the speaker voice characteristics rather than background noise, or sequence of words used to create the dataset.

\subsubsection{Dense Layers}
The 2 hidden dense layers before the output dense layers are crucial in the classification problem. They capture the necessary speach features extracted by the residual blocks and perform classification. A dropout of 0.2 is used in the first hidden layer which will be further detailed in Section.\ref{subsec:dropout}. The number of neurons in each dense layers, refer to Fig.\ref{fig:model_architecture_flow}, is tuned via trial and error on evaluation metrics that provide the best results, and most healthy accuracy and loss curves detailed in Section.\ref{sec:eval_metrics}.

\subsection{Model Regularization - Dropout}
\label{subsec:dropout}
We used dropout regularization in order to mitigate the model overfitting on certain characteristics of the person's speech. This 20\% dropout is tuned to ensure a maximal validation accuracy and steady convergence of validation loss. Without the presence of dropout, the validation loss curve becomes significantly erratic, and thus will not be able to generalize well on the speaker voice characteristics in the long run. This is detailed in Section.\ref{sec:eval_metrics}.

\subsection{Early Stopping Scheduler}
\label{subsec:es_scheduler}
It has been shown that training a model too much can have detrimental effects on the validation accuracy. If the amount of training and the validation accuracy is plotted, it shows a peak value and usually after further training, the model begins to overfit the training data, thus validation accuracy decreases. To ensure our validation accuracy is maximized for best performance, we keep a record of it (say $A_v$) for a number of training iterations, known as patience. 

In our case patience value was 10. That means if the validation accuracy does not increase during 10 epochs of training, the metrics at the epoch were $A_v$ was recorded is kept and training stops. This is called Early Stopping and is a commonly used practice in standard DL tasks.

\subsection{Learning Rate Scheduler}
\label{subsec:lr_scheduler}
In the beginning, a large learning rate is helpful as it allows the model to move quickly towards a point of convergence, which in our case, would indicate how well the model has learned to recognize voices.

In the later stages of learning, when the model is close to the point of convergence, small learning rates are desired so that the model does not overshoot and miss that point. This requires setting different learning rates at different stages of learning and is done with the learning rate scheduler.

In our model, the initial learning rate was set to 0.0001. After every 250 training steps, we scaled the learning rate down by 0.7. This is shown in Fig.\ref{fig:lr_curve}.

Hyperparameter tuning for learning rate was conducted with trial and error, and best performing learning rate with decay was saved according to results of model evaluations and metrics. The output biases and weights for the final dense layer conducting the multiple classification was checked to be converging to separate values. Moreover, the loss and accuracy curves were monitored to be healthy and converging without any abnormal behaviour, see Section.\ref{sec:eval_metrics}.

\newpage
\section{Evaluation Metrics}
\label{sec:eval_metrics}

\subsection{Model Optimization and Solution Convergence}
\label{subsec:optimization}
TensorBoard, a service from Tensorflow, was used to monitor model parameters.\cite{Tensorboard, UsingTensorboard} This allowed us to tune our model to our requirements and our type of dataset. 

\begin{figure}[ht!]
    \centering
    \includegraphics[width=1\linewidth]{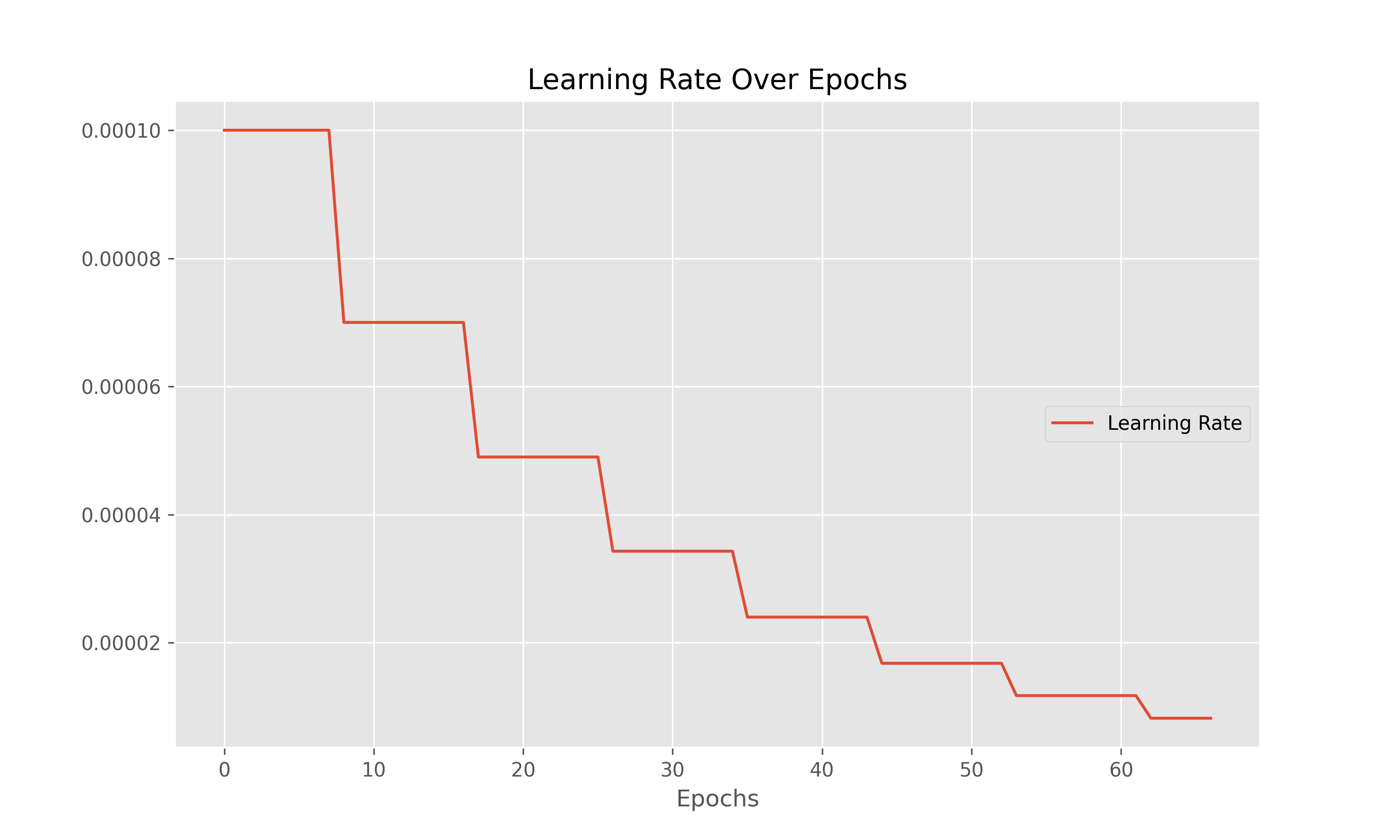}
    \caption{Learning rate scheduler working according to preset parameters described in Section.\ref{subsec:lr_scheduler}}
    \label{fig:lr_curve}
\end{figure}

Methods used to optimize and ensure model converged are:
\begin{enumerate}
    \item Reduce model complexity - We tuned parameters such as the residual block size, filter size and dense layer sizes
    \item Dropout regularization - so that training does not focus on unnecessary patterns in the data
    \item Learning rate schedule - hyperparameters in learning rate schedulers were tuned in order to achieve optimal curves, this ensures that higher learning rates towards the end of training do not hamper accuracy and loss. See Fig.\ref{fig:lr_curve}
    \item Increase batch size - often times, smooth accuracy curves exist but erratic validation loss curve may be an issue, which means that the validation set is too complex, or batch size of validation set used to evaluate the model was too small
\end{enumerate}

\subsection{Accuracy and Loss Curves}
During training, a close attention was kept on accuracy and loss curves. As the model trains, the accuracy of the model on both training set and and validation set will increase while the loss on each set will decrease. 

Occasionally in DL scenarios, even if the output predictions are accurate in practice, the model did not really converge to a point where maximum accuracy was achieved, but instead it stumbled upon a high accuracy point and held on-to that state due to EarlyStopping algorithm. This is where it is important to monitor the accuracy and loss curves to see that both training loss and validation loss are a smooth decreasing curve. And in addition, training accuracy and validation accuracy curves are smooth increasing curves.

Both curves should become asymptotic to a certain value after some epochs. And if performance deteriorates only then EarlyStopping must save the best state. See Section.\ref{subsec:es_scheduler} 

Validation accuracy should be slightly lower than validation training since the model will definitely be more accurate on the training set than the validation set. Similarly, validation loss must be slightly higher than training loss because of the aforementioned reason. \cite{ValLower}

In the base example code, this was not the case, as both accuracy and loss curves were fluctuating drastically, leading to uncertain training scheme. It was important to make some diagnostics and perform changes so that the model converged to a stable solution. 

\begin{figure}[ht!]
    \centering
    \includegraphics[width=1\linewidth]{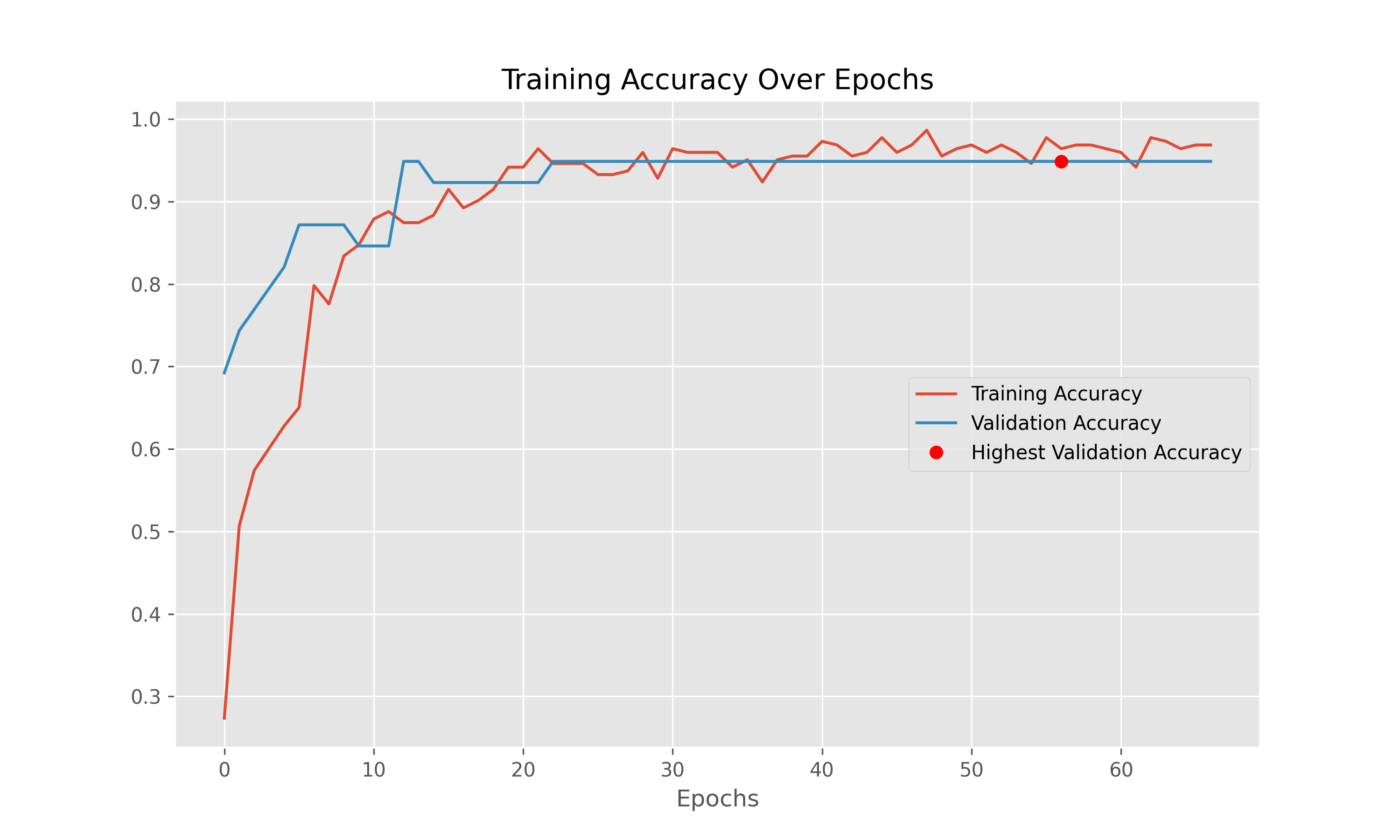}
    \caption{Plot of Training and Validation Accuracy versus number of Epochs, best checkpoint is shown in red, achieved through EarlyStopping}
    \label{fig:acc_curve}
\end{figure}

\begin{figure}[ht!]
    \centering
    \includegraphics[width=1\linewidth]{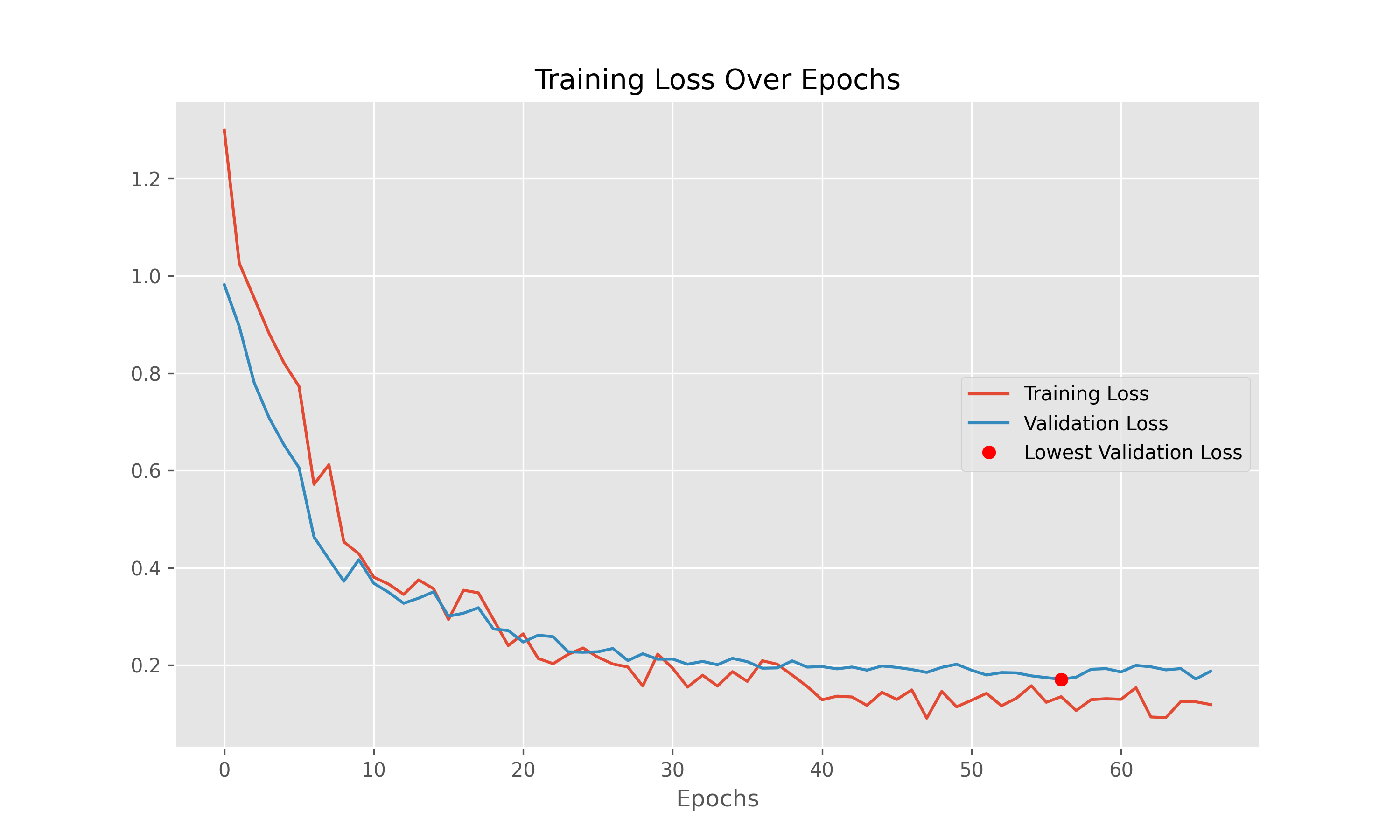}
    \caption{Plot of Training and Validation Loss versus number of Epochs, best checkpoint is shown in red, achieved through EarlyStopping}
    \label{fig:loss_curve}
\end{figure}

Our accuracy and loss metrics shown in Fig.\ref{fig:acc_curve} and Fig.\ref{fig:loss_curve} showcase that our model successfully converges and Earlystopping takes Epoch 56 parameters to save the model.

\newpage
\newpage
\subsection{Output Histograms}
Ensuring that our output layer histograms are sufficiently separated is a crucial observation in asserting that multiple classification is taking place. 

\begin{figure}[h]
    \centering
    \begin{subfigure}{0.5\linewidth}
        \includegraphics[width=\linewidth]{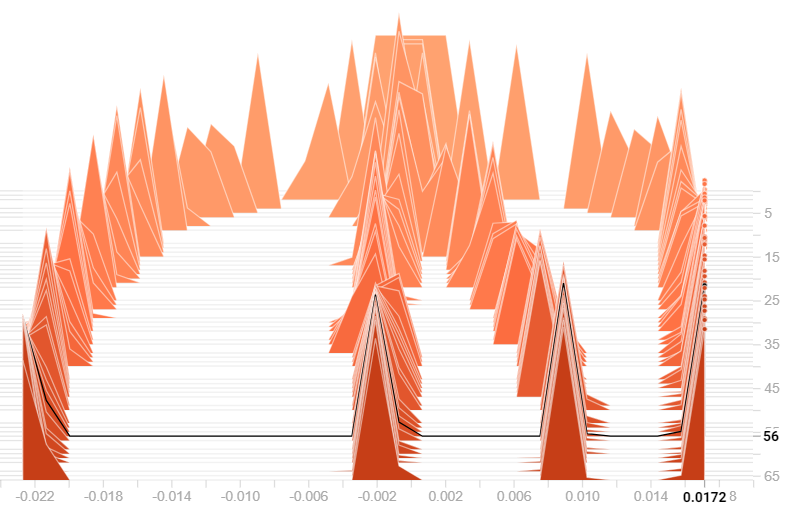}
        \caption{Bias parameters}
        \label{fig:output_bias}
    \end{subfigure}%
    \begin{subfigure}{0.5\linewidth}
        \includegraphics[width=\linewidth]{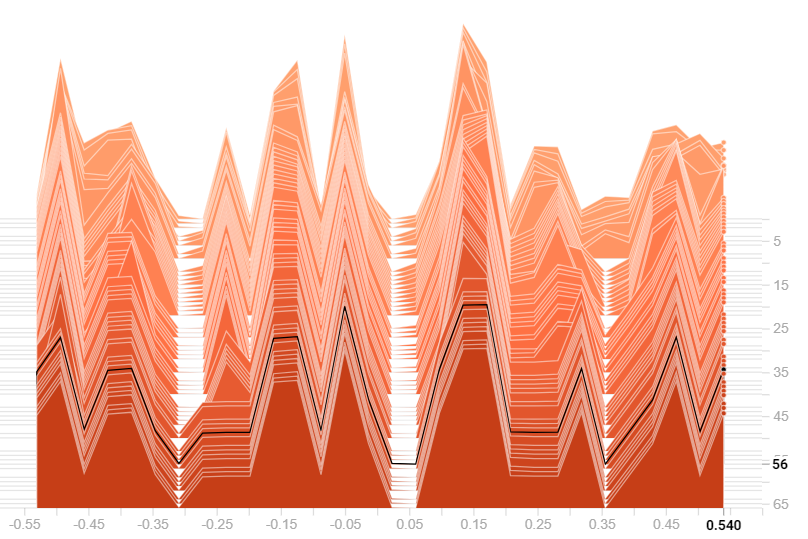}
        \caption{Weights parameters}
        \label{fig:output_weights}
    \end{subfigure}
    \caption{Output Dense Layer Parameter Histograms. EarlyStopping point is indicated at epoch 56.}
    \label{fig:output_params}
\end{figure}

We can see from Fig.\ref{fig:output_bias} that the bias values of the (by default 4 classes) classes are separating into their respective values. Moreover we can observe that there is no direction changing in out bias values amidst training. A change of direction of the bias histogram as the model trains would mean that the model is overshooting the convergence point and not learning effectively. This was addressed using the steps mentioned in Section.\ref{subsec:optimization}. 

In Fig.\ref{fig:output_weights} is it seen that the weights of each neurons are separated. A histogram that is distributed in fluctuating peaks throughout a large range about x=0 values indicated that the model is learning positive and negative features equally and the multiple peaks ensure that the model is learning specific features within the data.

\newpage
\section{Results}
\label{sec:results}
From our experimentation, we were able to optimize our model to detect and identify speech features within a voice clip as small as 8 seconds in length using comparatively simple 1D-CNN model. Training of the model was complete within on average 1 minute on a GPU Nvidia RTX 3060 4GB on an AMD Ryzen-7 5800H 16 Core CPU. Moreover the model successfully converged in low number of epochs, and therefore low number of iterations.

Moreover, hyperparameters such as learning rate, layer sizes, batch sizes and dropout were tuned to attain the optimal performance.

Our model was successful in identifying speakers in our dataset with a validation accuracy 97.87\% even with added random noise. This means that on cleaner voice inputs the accuracy of prediction should be larger. Even with a considerably small dataset, the model was tuned to perform significantly well on unseen data. 

A GUI was created to make model usage and distribution easier and simpler.

Future experiments will evaluate model performance on large datasets like VoxCeleb1 and LibriSpeech to validate scalability. Additionally, pretrained models such as wav2vec will be tested as baselines for comparison.

\section{Conclusion}
\label{sec:conclusion}
This paper demonstrates the feasibility of speaker identification on minimal datasets using 1D-CNNs, achieving competitive validation accuracy. Key contributions include the implementation of augmentation techniques and the optimization of lightweight architectures. There are a few different future improvements that can be made. Firstly, the model can be further tuned using programmable tuners available in tensorflow. A comparative study can be done on hyperparamater changes on the model performance using Tensorboard. Residual blocks can be modified with difference architecture, adding BatchNormalization layers that might enable a better performance on the multiple classification task, such as done in ResNet DNN models. Furthermore, hyperparameter tuning for L2 Regularization can be done in the future such as for weight decay algorithm. Different kinds of initialization schemes for each layer can be used to observe the differnce on the model training and performance. Future directions involve testing on larger, more diverse datasets and leveraging transfer learning to improve model scalability.

\newpage
\bibliography{bibs}






\end{document}